\documentstyle[preprint,aps,eqsecnum]{revtex}
\begin{document}
\draft
\title
{A NONPERTURBATIVE CALCULATION OF BASIC  \\
 CHIRAL QCD PARAMETERS WITHIN ZERO MODES \\
 ENHANCEMENT MODEL OF THE QCD VACUUM. I }

\author{V. Gogohia }

\address{ RMKI, Department of Theoretical Physics,
Central Research Institute for Physics, \\
H-1525,  Budapest 114,  P. O. B.  49,  Hungary}

\maketitle

\begin{abstract}
A new zero modes enhancement (ZME) model of the true QCD vacuum
is breifly described. It makes possible to
analytically investigate and calculate numerically
low-energy QCD structure from first principles.
Expressions of basic
chiral QCD parameters (the pion decay constant, the quark and
gluon
condensates, the dynamically generated quark mass, etc) as well as
the vacuum energy density (up to the sign, by definition, the bag
constant), suitable for numerical calculations, have been derived.
Solution to the Schwinger-Dyson
(SD) equation for the quark propagator in the infrared (IR) domain
on the basis of the ZME effect in QCD was used for this
purpose. There are only two independent quantities (free
parameters) by means of which  calculations must be done
within our approach. The first one is the integration constant
of the above mentioned quark SD equation of motion. The second one
is a scale at which nonperturbative effects begin to play a dominant role.
\end{abstract}

\pacs{ PACS numbers: 11.30 Rd, 12.38.-t, 12.38 Lg and 13.20 Cz.}

\vfill
\eject

\section{The zero modes enhacement model of quark confinement
            and DCSB}

    Today there are no doubts left that the dynamical mechanisms
of quark confinement [1] and dynamical chiral symmetry breaking
(DCSB) [2, 3]
are closely related to the complicated topological structure
of the QCD nonperturbative vacuum. For this reason
any correct nonperturbative model of quark confinement and
DCSB necessary becomes a model of a true QCD vacuum and vice versa.
Also, it becomes clear that nonperturbative infrared (IR)
divergenses are closely related
to the above mentioned nontrivial vacuum structure,
play an important role in the large scale
behaviour of QCD [4-6]. If it is true that QCD is an IR
unstable theory (i.e., if has no IR stable fixed point) then the
low-frequency modes of
the Yang-Mills fields should be enhanced due to nonperturbative IR
divergences. So the full gluon propagator can diverge faster than
the free one at small momentum, in accordance with [7-10]
\begin{equation}
D_{\mu\nu}(q) \sim (q^2)^{-2} ,
\qquad q^2 \rightarrow 0
\end{equation}
which describes
the zero modes enhancement (ZME) effect in QCD.
If, indeed the low-frequency components of the virtual fields
in the true vacuum have a larger amplitude than those of the bare
(perturbative) vacuum [5], then the Green's function  for a single
quark should be reconstructed on the basis of this effect. It is
important to understand that a
possible effect of the ZME (1.1) is our
primary dynamical assumption. We will consider this effect as
a rather realistic confining ansatz for the full gluon propagator
in order to use it as input information for the quark SD equation.

   Let us consider the exact, unrenormalized
SD equation for the quark propagator in momentum space (Fig. 1)
\begin{equation}
S^{-1}(p) = S^{-1}_0(p) + g^2 C_F \int {d^nq\over {(2\pi)^n}}
\Gamma_\mu(p, q) S(p-q)\gamma_\nu D_{\mu\nu}(q),
\end{equation}
where  $C_F$  is the eigenvalue of the quadratic Casimir
operator in the fundamental representation. Since the notions are
obvious, let us only note that our parametrization of
the full quark propagator is as follows
\begin{equation}
         -iS(p)= \hat pA(-p^2)+B(-p^2).
\end{equation}
The full gluon
propagator in the arbitrary covariant gauge is
\begin{equation}
D_{\mu\nu}(q) = - i \left\{ \left[ g_{\mu\nu} -
{{q_\mu q_\nu}\over {q^2}} \right]
{1\over {q^2}} d(-q^2,  a) + a {{q_\mu q_\nu}\over
{q^4}} \right\},
\end{equation}
where $a$ is a gauge fixing parameter ($a = 0$ for Landau gauge).

  Assuming that in the IR region
\begin{equation}
d(-q^2,  a) = \left( {{\mu^2}\over {-q^2}} \right) +
\beta(a) + O(q^2),
\qquad q^2 \rightarrow 0,
\end{equation}
where $\mu$ is the appropriate  mass scale parameter,  we obtain
the above mentioned generally accepted form of the IR singular
asymptotics for the full gluon propagator (1.1) [4, 5, 7-10]
(enhancement of the zero modes).

    In order to actually define an initial SD equation
(1.2) in the IR region (at small momenta) let us
apply the gauge-invariant dimensional regularization
method of 't Hooft and Veltman [11] in the limit $n = 4 + 2
\epsilon$, where $\epsilon$ is
a small IR regulation parameter, which
is to be set to zero at the end of the computation
$\epsilon \rightarrow 0^+$. Because of this the
quark propagator and other Green's functions become dependent, in
general, on this IR regulation parameter $\epsilon$. We consider
the SD equations and the corresponding quark-gluon ST
identity in Euclidean space
($d^nq \rightarrow i d^nq_E,  \quad q^2  \rightarrow - q^2_E,
\quad p^2  \rightarrow -p^2_E,  \quad $
but for simplicity the Euclidean subscript  will be omitted).

 Let us use, in the sense of the distribution theory, the relation
[12]
\begin{equation}
(q^2)^{-2+\epsilon} = {{\pi^2}\over {\epsilon}} \delta^4(q) +
(q^2)^{-2}_+ + O(\epsilon), \qquad \epsilon \rightarrow 0^+,
\end{equation}
where $(q^2)^{-2}_+$ is a functional acting on the
main (test) functions according to the so-called "plus
prescription" standard  formulae [12].
   Substituting (1.4-1.6) into the quark SD equation (1.2) on
account of the above mentioned "plus prescription" formulae and
expanding in powers of $q$ and keeping the terms of order
$q^{-2}$ ( the Coulomb order terms), in agreement with (1.5),
one finally arrives at the quark propagator
expansion in the IR region for unrenormalized quantities [13]
(in four-dimensional Euclidean space).

   As mentioned above, all Green's functions
become dependent generally on the IR regularization
parameter $\epsilon$.
In order to extract the finite Green's functions in
the IR region, we introduce the renormalized
(IR finite) quark-gluon vertex function at zero momentum
transfer and the quark propagator as follows
$\Gamma_\mu(p, 0) = Z_1(\epsilon) \bar \Gamma_\mu(p, 0)$
and
$S(p) = Z_2(\epsilon) \bar S(p)$ at $\epsilon \rightarrow  0^+$,
respectively. Here $Z_i(\epsilon)\ (i=1,2)$
are the corresponding IR renormalization constants.
The $\epsilon$-parameter dependence is indicated explicitly
to distinguish them from the usual UV renormalization
constants.
$\bar \Gamma_\mu(p, 0)$ and $\bar S(p)$ are the renormalized
(IR finite) Green's functions and therefore do not depend
on $\epsilon$ in the
$ \epsilon \rightarrow  0^+ $ limit, i.e. they exist as
$ \epsilon \rightarrow  0^+ $.
The correct treatment of such strong singularity (1.6)
within distribution theory [12]
enabled us to extract the required class of test functions
 in the renormalized quark SD equation. The test functions do
 consist of the quark propagator and the corresponding quark-gluon
vertex function. By the renormalization program we have
found the regular (at zero point) solutions for the quark
propagator (see
below). For that very reason relation (1.6) is justified, it is
multiplied by the appropriate smooth test functions.
  Due to a quark convergence condition, a cancellation of
nonperturbative IR divergences takes place. Because of this
condition the explicitly gauge-dependent terms (the so-called
next-to-leading
terms) in the above mentioned quark propagator expansion
become $\epsilon$ - order
 terms. For this reason these noninvariant terms vanish in the
$\epsilon \rightarrow  0^+ $ limit [13].

  Absolutely in the same way should be reconstructed the ghost
self-energy and the corresponding ST identity for the quark-gluon
vertex [13]. We develop a method for the extraction of the
IR-finite Green's functions in QCD which
means that they do not depend on the
IR regulation parameter $\epsilon$ as $\epsilon \rightarrow 0^+$.
For this purpose we have worked out a renormalization program
in order to cancel all the IR nonperturbative
divergences which makes it possible to obtain a close set
of the SD equations and the corresponding ST identity in the
quark sector.
By completing our renormalization program, we explicitly show
that only multiplication by the quark IR renormalization
constant will remove all nonperturbative
IR divergences from the theory on a general ground.
We have shown [13]
that for the covariant gauges the complications due to
ghost contributions can be considerable in our approach.

 The closed set of equations, obtained in
our paper [13], which will be used for numerical calculation
of basic chiral QCD parameters as well as the vacuum energy
density, should read
\begin{equation}
S^{-1}(p)= S_0^{-1} (p) + \tilde g^2 \bar \Gamma_\mu(p,0)
S(p) \gamma_\mu,
\end{equation}
\begin{equation}
{1\over 2}\bar b(0) \bar  \Gamma_\mu(p,0) =
i \partial_{\mu} S^{-1}(p) - {1\over 2}\bar b(0) S(p)
 \bar \Gamma_\mu(p,0) S^{-1}(p),
\end{equation}
where $S^{-1}_0(p)$
is the free quark propagator. The finite at zero point the ghost
self-energy is denoted by $\bar b(0)$ and $\tilde g^2$ includes
the mass scale parameter $\mu^2$, determining the validity of the
above mentioned deep IR singular asymptotic behaviour
of the full gluon propagator (1.5).
Let us mention that the IR finite quark renormalization constant,
explicitly not shown here,
which should multiply the free quark propagator in Eq. (1.7) is
to be set to unity without losing generality (multiplicative
renormalizability) [13]. It is worth noting also that Eq. (1.7)
and Eq. (1.8) describe the leading terms of the corresponding
expansions of the quark SD equation and ST identity in the
IR region, respectively [13].

     In order to solve the system (1.7-1.8), it is
 convenient to represent the quark-gluon vertex
 function at zero momentum transfer as follows
\begin{equation}
\bar \Gamma_\mu(p,0) = F_1(p^2) \gamma_\mu + F_2(p^2)p_\mu
 + F_3(p^2)p_\mu \hat p + F_4(p^2) \hat p\gamma_\mu.
\end{equation}
  Substituting this
 representation into the ST identity (1.8), one obtains
\begin{eqnarray}
 F_1(p^2)&=& - \overline A(p^2), \nonumber \\
 F_2(p^2)&=& - 2 \overline B'(p^2) - F_4(p^2),  \nonumber\\
 F_3(p^2)&=&  2 \overline A'(p^2),    \nonumber\\
 F_4(p^2)&=&  {{A^2(p^2)B^{-1}(p^2)} \over {E(p^2)}}.
\end{eqnarray}
 Here the prime denotes differentiation with respect to the
 Euclidean momentum variable  $p^2$ and
$\overline A(p^2) = A(p^2)E^{-1}(p^2)$,
$\overline B(p^2) = B(p^2)E^{-1}(p^2)$ with
$E(p^2) = p^2 A^2(p^2) + B^2(p^2)$.
For the sake of convenience, the ghost self-energy at
zero point $\bar b \equiv \bar b(0)$ is included into
the definition of a new coupling constant
$\lambda = g^2[\overline b(0)]^{-1}(2\pi)^{-2}$.

   Proceeding now to the dimensionless variables by
$p^2 = \mu^2 t = \mu^2 { \lambda \over 2} z$ and parameters
${2 \over \lambda } t_0 = z_0, t_0 = { k^2_0 \over \mu^2}$,
and introducing then the following notations
\begin{equation}
A(p^2) = \mu^{-2}A(t) = - \mu^{-2} {2 \over \lambda} g(z), \
B^2(p^2) = \mu^{-2}B^2(t) = \mu^{-2} {2 \over \lambda} B^2(z_0, z),
\end{equation}
and doing some algebra, the initial system (1.7-1.8)
can be  rewritten as follows (normal form)
\begin{equation}
g'(z)= - [{2 \over z} + 1] g(z) + {1 \over z} + \tilde{m_0}B(z)
\end{equation}
\begin{equation}
B'(z) = -{3\over 2} g^2(z) B^{-1}(z) - [\tilde{m_0} g(z) + B(z)],
\end{equation}
where $\tilde{m_0} = m_0({2 \over \lambda})^{1/2}$. It is easy to
check that solutions to this system in the chiral limit $m_0=0$
are:
\begin{equation}
g(z) =  z^{-2} [\exp{(-z)} -1 + z]
\end{equation}
and
\begin{equation}
B^2(z_0, z) =  3 \exp{(-2z)}
\int \limits_z^{z_0} {\exp{(2z')} g^2(z')\,dz'}.
\end{equation}

The  exact solutions (1.14) for $g(z)$ and (1.15) for the
dynamically generated quark mass function $B(z_0, z)$
are not entire functions. The functions $g(z)$
and $B(z_0, z)$ have removable singularities at zero. In
addition, the dynamically generated quark mass function
$B(z_0, z)$ also has algebraic branch points at $z=z_0$ and
at infinity. Apparently, these
unphysical singularities are due to ghost contributions.
The quark propagator may or may not be an entire function
but in either
cases the pole-type singularities should disappear. This is a
general feature of quark confinement and holds in any gauge.

 In order to reproduce automatically the
correct behaviour of the dynamically generated quark
mass function at infinity, it is necessary to put $z_0=\infty$ in (1.15)
from the very beginning. Obviously, in this case solution (1.15)
cannot be accepted at zero $z=0$, so one needs
to keep the constant of integration $z_0$ arbitrary but finite
in order to obtain a regular, finite solution at zero point.

  The region $z_0 > z$ can be considered as nonperturbative,
whereas the
region $z_0 \le z$ can be considered as perturbative.
By approximating the full gluon propagator by its
deep IR asymptotics such as  $(q^2)^{-2}$ in the whole
range
$\left[ 0,  \infty \right) $,  we nevertheless obtain that our
solution  for the dynamical quark mass function $B(z_0, z)$
manifests
the existence of the boundary value momentum (dimensionless)
$z_0$ which separates the IR (nonperturbative) region
from the intermediate and UV (perturbative) regions.
If QCD confines then a characteristic scale, at which
confinement and other nonperturbative
effects become essential, must exist.
On the other hand, because of this one can eliminate
the influence of the above mentioned unphysical singularities,
coming from
the solutions to the quark SD equations (due to necessary ghost
contributions), on the $S$-matrix elements reproducing physical
quantities.

  Thus, within our approach to QCD at large
distances [13] in order to obtain numerical values
of any physical quantity, e.g. the pion decay
constant (see below), the integration over the whole range
$\left[ 0, \  \infty \right] $
reduces to the integration over the nonperturbative region
$ \left[ 0, \  z_0 \right], $  which determines the range of
validity of the deep IR asymptotics (1.1) of the full gluon
propagator and consequently the range of validity of the
corresponding solutions (1.14) and (1.15) for the IR piece of
the full quark propagator.

  Let us make the main conclusions now. First, if the
enhancement of the zero modes of the vacuum
fluctuations (1.1) takes place indeed then the quark Green's
function, reconstructed on the basis of this effect, has no poles.
In other words, the enhancement of the zero modes at the
expense of the virtual gluons alone removes single
quark from the mass-shell (quark confinement theorem of Ref. 13).
Second, a chiral symmetry violating part of the quark
propagator in this case is automatically generated. From the
obtained system (1.12-1.13) it explicitly follows that a chiral
symmetry preserving solution ($m_0 = B(z) = 0,  g(z) \ne 0 $)
$ is \ forbidden$. So a chiral symmetry violating
solution ($m_0 = 0, B(z) \ne 0,  g(z) \ne 0 $) for the quark
SD equation $is \ required$. Thus the enhancement of the zero
modes automatically leads to quark confinement and DCSB at the
fundamental quark level and they are in close connection with each other.

\section{The vacuum energy density and gluon condensate }

  The effective potential method for composite operators [14]
enables us to investigate the vacuum of QCD since, in the
absence of the external sources, the effective potential is
nothing
but the vacuum energy density, the main characteristic of the
vacuum. The effective potential at one-loop level is [14]
\begin{eqnarray}
V(S, D) &=& V(S) + V(D) =  \nonumber\\
&-& i \int {d^np \over {(2\pi)^n}}
Tr\{ \ln (S_0^{-1}S) - (S_0^{-1}S) + 1 \} \nonumber\\
&+& i { 1 \over 2} \int {d^np \over {(2\pi)^n}}
 Tr\{ \ln (D_0^{-1}D) - (D_0^{-1}D) + 1 \},
\end{eqnarray}
where $S(p)$ (1.3), $S_0(p)$ and $D(p)$ (1.4), $D_0(p)$
are the full, free quark and gluon propagators, respectively.
Here and everywhere below the trace over space-time
and color group indices is assumed but they are suppressed.
Let us recall that the free gluon propagator can be
obtained from (1.4) by setting simply $d(-q^2, a) = 1$.

 Evidently the effective potential is normalized as follows
$V(S_0, D_0) = V(S_0) = V(D_0) = 0$. Because of this normalization
the vacuum energy density now should be defined as follows
$\epsilon = V(S_0, D_0) - V(S, D) = - V(S,D)$. This means that the
perturbative vacuum is normalized to zero and
$\epsilon = \epsilon_q + \epsilon_g$
with $\epsilon_q = - V(S), \ \epsilon_g = - V(D)$,
where $V(S)$ and $V(D)$ are given by (2.1).

  Going over to Euclidean space
($d^4p \rightarrow i d^np, \quad p^2 \rightarrow -p^2$ )
and dimensionless variables and parameters (1.11), we finally
obtain after some algebra ($n=4$)
\begin{equation}
\epsilon_q = - {3 \over 8 \pi^2} k^4_0 z^{-2}_0
\int \limits_0^{z_0} dz\, z\, \{ \ln z\left[ z g^2(z) +
B^2(z_0, z)\right] - 2z g(z) + 2\},
\end{equation}
where we introduced the UV cutoff which should be
identified with the arbitrary constant
of integration $z_0$ as was discussed in previous section.
The explicit expressions for $g(z)$ and $B^2(z_0, z)$ are given
by (1.14) and (1.15) respectively. The expression (2.2) describes
the comtributions to the vacuum energy density of the light
confining quarks wuth dynamically generated masses.

The vacuum energy density due to
the nonperturbative gluon contributions in the same variables is
\begin{equation}
\epsilon_g =  {1 \over \pi^2} k^4_0 z^{-2}_0
\int \limits_0^{z_0} dz\, z\, \{ \ln (1+ {6 \over z})
 - {3 \over 2z } + b\}
\end{equation}
Here one important remark is in order. In fact, vacuum energy
density
$\epsilon_g$ does not vanish at $z_0 \rightarrow \infty$ as it
should because of the above mentioned normalization.
Thus it needs an additional
regularization at this limit. From (2.3), it follows that the
term containing the constant $b$ should be substracted from this
expression. So regularized vacuum energy density should be
calculated through the relation (2.3) which becomes
\begin{equation}
\epsilon_g = - {1 \over \pi^2} k^4_0 z^{-2}_0 \times
\left[ 18 \ln (1 + { z_0 \over 6}) - {1 \over 2} z^2_0
\ln (1 + {6 \over z_0}) - {3 \over 2} z_0 \right].
\end{equation}
The vacuum energy density due to confining quarks
(2.2) automatically disappears at $z_0 \rightarrow \infty$, so it
does not requires any additional regularization.

The vacuum energy density
is important on its own right as the main characteristics of the
nonperturbative vacuum of QCD. On the other hand, it makes
possible to estimate such important phenomenogical parameter as
the gluon condensate introduced within the QCD sum rules approach
to resonance physics [15]. Indeed, through the vacuum energy
density $\epsilon$ it can be expressed as follows
\begin{equation}
\langle{0} | {\alpha_s \over \pi} G^a_{\mu\nu} G^a_{\mu\nu} |
{0}\rangle = - {32 \over 9} \epsilon
      = - {32 \over 9} (\epsilon_q + \epsilon_g).
\end{equation}
The weakness of
this derivation [15] is, of course, that it was obtained on the
basis
of the perturbative calculation of the $\beta(\alpha_s)$-function.
In any case, it would be instructive to estimate the gluon
condensate with the help of (2.5).

\section{ Basic chiral QCD parameters }

Because of its especially
small mass, the pion is the most striking example of the Goldstone
realization of chiral symmetry $SU(2)_L \times SU(2)_R$.
Beside the quark condensate and the dynamically
generated quark mass it is one of the three important chiral
QCD parameters that determine the scale of chiral dynamics.

I. The pion
decay constant $F_{\pi}$ is defined in the current algebra (CA) as
\begin{equation}
\langle{0}| J^i_{5\mu}(0) |{\pi^j(q)} \rangle = iF_{\pi}q_\mu
\delta^{ij} .
\end{equation}
(The normalization $F_{\pi} = 92.42 \ MeV$ is used [16]).
Clearly, this matrix element can be written in terms of the pion-
quark-antiquark proper vertex and quark propagators as
\begin{equation}
 iF_{\pi}q_{\mu} \delta^{ij} = \int {d^4p\over {(2\pi)^4}} Tr\bigl\{
 \left({\lambda^i \over 2}\right)\gamma_5\gamma_{\mu}
S(p+q)G^j_5(p+q,p)S(p)\bigr\}.
\end{equation}
To get expression for $F_{\pi}$  one has to differentiate
Eq. (3.2) with respect to the external momentum
$q_{\nu}$  and then set $q=0$ (Fig. 2).

   Information on the BS pion wave function up
to terms of order $q$ can be obtained from the corresponding
axial-vector vertex.
Indeed, in our paper [17] it has been found that this vertex can
be decomposed in a self-consistent way into pole (dynamical)
and regular parts as follows
\begin{equation}
\Gamma^i_{5\mu}(p+q,p) =
- {q_\mu \over {q^2}} G_5^i(p+q,p)
+ \Gamma^{iR}_{5\mu}(p+q,p),
\end{equation}
where the BS bound-state amplitude is
\begin{equation}
G_5^i(p+q,p)= - {1\over{F_{\pi}}}\left({\lambda^i \over
2}\right)\gamma_5 G(p+q,p),
\end{equation}
with
\begin{equation}
G(p+q,p) = 2 \overline B(-p^2)+\hat q R_6(-p^2)+\hat p \hat q R_{11}(-p^2)
\end{equation}
and arbitrary form factors are the residues of the corresponding
form factors, entering the vertex from the very beginning [17].
  The regular part at zero momentum transfer $q=0$ is determined
as
\begin{equation}
\Gamma^{iR}_{5\mu}(p,p) = \bigl( {\lambda^i\over 2}\bigr)
\gamma_5 \{ \gamma_\mu G_1 + p_\mu G_2 + p_\mu \hat p G_3
+ \hat p \gamma_\mu G_4  \},
\end{equation}
where in the Euclidean space
\begin{eqnarray}
G_1(p^2) &=& - \overline A (p^2) - R_6(p^2)  \nonumber\\
G_2(p^2) &=&  2 \overline B'(p^2) \nonumber\\
G_3(p^2) &=&  2 \overline A'(p^2) \nonumber\\
G_4(p^2) &=& - R_{11}(p^2).
\end{eqnarray}
This system is nothing else but the conditions for the
cancellation of the dynamical singularities at $q=0$ for the
corresponding form factors [17]. The
regular part at $q=0$ also depends on the same form factors
$R_6$ and $R_{11}$ as the BS bound-state amplitude up to terms
of order $q$.

   Then by taking into account the BS pion wave function up
to terms of order $q$, as given by (3.4-3.5), and (1.3) with the
substitution $p \rightarrow p+q$ and expanding in powers of $q$,
the expression (3.2) can easily be evaluated.
Going over to Euclidean space
$(d^4 p \rightarrow i d^4 p,\ p^2 \rightarrow - p^2)$
and using dimensional variables (1.11), one finally
obtains
\begin{eqnarray}
F^2_{CA} = {{12\pi^2}\over{(2\pi)^4}}\mu^2
   {\int}^{\infty}_0 dt\, t \{ -\overline B(t)
   [AB+{1\over 2}t(A'B-AB')] \nonumber\\
   - {3\over 4}tABR_{11}(t) + {1\over 4}R_6(t)(E-3B^2) \},
\end{eqnarray}
Here the primes denote differentiation with
respect to the dimensionless Euclidean momentum variable $t$,
$A=A(t), B=B(t)$ and quantities with overline are shown after
Eqs. (1.10). Here and in what follows we denote the pion
decay constant $F_{\pi}$ in the chiral limit of the CA
representation by $F_{CA}$.

   The main problem now is to find a good
nonperturbative ansatz for both arbitrary
form factors $R_j(t) (j=6,11)$ in the IR region, which determine
the contributions to the pion decay constant from the second
diagram in Fig. 2.
 In the nonperturbative calculations these terms cannot be ignored
by saying formally they are of order $g^2$ in the coupling
constant, as it was done in the perturbative treatments [3,
18-19]. In connection with this let us point out
that the difference between the vector and axial-vector currents
disappears in the chiral limit.  For this reason let us assume
[20] that the IR finite quark-gluon vector vertex function at
zero momentum transfer (1.9-1.10) is
a good approximation to the regular piece of the axial-vector
vertex at zero momentum transfer in the chiral limit (3.6-3.7).
A fortunate feature that admits to exploit partial analogy
between
vector and axial-vector currents in the chiral limit for the
flavor non-singlet channel is that the contribution
to the pion decay constant in the CA representation (3.8)
does not depend on the
form factor $G_2(p^2)$ at all.  In this case the analogy between
(1.10) and (3.7) becomes complete, and one obtains
\begin{equation}
R_6(p^2) = 0, \qquad
R_{11}(p^2) = - {{A^2(p^2)B^{-1}(p^2)} \over {E(p^2)}}.
\end{equation}
Of course, we cannot prove  these relations but it will be
shown later (Part II) that this dynamical assumption
(nonperturbative
ansatz) leads to very good numerical results for all chiral QCD
parameters thereby justifying it once more.

In terms of the new parameters and variables (1.11) and on account
of (3.9) we finally recast (3.8) as follows
\begin{equation}
F^2_{CA} = {3\over {8\pi^2}} k_0^2 z_0^{-1}
             \int^{z_0}_0 dz \,{ zB^2(z_0,z) \over
             {\{zg^2(z) + B^2(z_0,z)\}}}.
\end{equation}
This expression will be used for numerical
calculation of the pion decay constant in the CA representation.

 II. The other important chiral QCD parameter is the dynamically
generated quark mass $m_d$,
 defined as the inverse of the full quark
propagator (1.3) in the chiral limit at zero point [13, 17, 21]
\begin{equation}
m_d = [iS_{ch}(0)]^{-1},
\end{equation}
where $S_{ch}(0)$ denotes the full quark propagator in the chiral
limit $m_0 =0$.
Obviously, this definition assumes
also regularity at the zero point. Though the dynamical
quark mass $m_d$ is not an experimentally observable
quantity, by all means, it is desirable to find
such kind of solutions to the quark SD equations
in which dependence on a gauge-fixing parameter
disappears. In this sense $m_d$ defined by
(3.11) becomes gauge-invariant. As it was
briefly discussed in Section 1, exactly such a
nonperturbative quark propagator has been found within
our approach to QCD at large distances [13].

Using the standard decomposition of the quark propagator (1.3)
and its inverse, dynamical chiral symmetry breaking (DCSB) at the
fundamental (microscopic)
quark level can be implemented by the following condition
\begin{equation}
\bigl\{ S^{-1}(p),  \gamma_5 \bigr \}_+ = i \gamma_5
2 \overline B(-p^2) \ne 0,
\end{equation}
so that the $\gamma_5$ invariance of the quark propagator is
broken
and the measure of this breakdown is the double of the dynamically
generated quark mass function  $2 \overline B (-p^2)$.
Precisely this quantity at zero $2 \overline B(0)$ can be defined
as a scale of DCSB at the fundamental quark level [17].
In accordance with (3.11), let us denote it by
\begin{equation}
 \Lambda_{CSBq} = 2 \overline B(0) = 2 m_d ,
\end{equation}
The definitions
$m_d$ and $\Lambda_{CSBq}$ have now direct physical
sense within the above mentioned solutions to the quark SD equation.

  Let us write down the final result for the
dynamically generated nonperturbative quark mass (3.11) too,
expressed in terms of the new parameters and variables (1.11)
\begin{equation}
m_d = k_0\bigl\{z_0 B^2(z_0,0)\bigr\}^{-1/2},
\end{equation}
where $B^2(z_0,0)$ is given by (1.15) at zero point.

 III. As it is well known, the order parameter of DCSB - quark
condensate
can also be expressed in terms of the quark propagator scalar
function $B(-p^2)$ (1.3). Its definition is
\begin{equation}
{\langle \overline qq \rangle} =
{\langle 0| \overline qq | 0 \rangle}
=  \int {d^4p\over {(2\pi)^4}} Tr S(p).
\end{equation}
The final result expressed
in terms of new variables and parameters (1.11) is as follows
\begin{equation}
{\langle \overline qq \rangle}_0 = -
{3\over {4\pi^2}}k_0^3z_0^{-3/2}{\int}^{z_0}_0 dz\,{zB(z_0,z)},
\end{equation}
wher the superscript $0$ denotes the chiral limit. As a
function of $m_d$ the quark condensate can be expressed on
account of (3.14).

Thus there are only two independent (free)
quantities by means of which all calculations should be done
in our approach. The first one is the constant of integration
$z_0$ of the above mentioned quark SD equation of motion.
The second quantity is a scale $k_0$ at
which nonperturbative effects begin to play a dominant role.
  The ZME model of a true QCD vacuum enables us to describe
quark confinement and DCSB on a general ground. We begin Part II
with numerical investigation of the low energy QCD structure
at the chiral limit. At low energies QCD is under control of
$SU_L(N_f) \times SU_R(N_f)$ chiral symmetry ($N_f$ is the
number of different flavors) and its dynamical breakdown in the
vacuum to the corresponding vectorial subgroup. So to correctly
calculate basic low energy QCD parameters in the chiral limit
means to corectly understand the dynamical structure of QCD at low
energies. That is why it is important to start from
the chiral limit.

\vfill
\eject

 \vfill
 \eject

\begin{figure}

\caption{The SD equation for the quark propagator in momentum
space.  }

\bigskip

\caption{ An exact expression for the pion decay constant
$F_{\pi}$ in the current algebra (CA) representation.
Here $G^j_5$, $S$ and $J^i_{5\mu}$ are the pion-quark bound-state wave
function, the quark propagator and the  axial-vector current, respectively.
The slash denotes differentiation with respect
to momentum $q_{\nu}$  and setting $q = 0$.}
\end{figure}

\end{document}